\documentclass[twocolumn,showpacs,preprintnumbers,amsmath,amssymb,prb]{revtex4}

\usepackage{graphicx}
\usepackage{dcolumn}
\usepackage{bm}

\usepackage{color}

\newcommand{\ket}[1]{| #1 \rangle}

\begin{document}

\preprint{APS/123-QED}

\title{Easy-plane magnetocrystalline anisotropy in the multi-step metamagnet CeIr$_3$Si$_2$}

\author{K. Shigetoh$^{1}$, A. Ishida$^{1}$, Y. Ayabe$^{1}$, T. Onimaru$^{1}$, K. Umeo$^{2}$, Y. Muro$^{4}$, K. Motoya$^{4}$, M. Sera$^{1}$, and T. Takabatake$^{1,3}$}
\affiliation{%
$^{1}$Department of Quantum Matter, ADSM, $^{2}$Materials Science Center, N-BARD,\\
$^{3}$Institute for Advanced Materials Research, Hiroshima University, Higashi-Hiroshima 739-8530, Japan\\
$^{4}$Department of Physics, Faculty of Science and Technology, Tokyo University of Science, Noda 278-8510, Japan}

\date{\today}

\begin{abstract}
Highly anisotropic properties of CeIr$_3$Si$_2$ have been observed by the magnetization $M$($B$), electrical resistivity $\rho$, and specific heat measurements on a single-crystalline sample. This compound with an orthorhombic structure having zigzag chains of Ce ions along the $a$-axis undergos magnetic transitions at 3.9 K and 3.1 K. At 0.3 K, metamagnetic transitions occur at 0.68 T and 1.3 T for $B$$//$$b$ and 0.75 T for $B$$//$$c$. Easy-plane magnetocrystalline anisotropy is manifested as $M$($B//b$) $\cong$ $M$($B//c$) $\cong$ 11$M$($B//a$) at $B$ = 5 T. Electrical resistivity is also anisotropic; $\rho_{b}$ $\cong$ $\rho_{c}$ $\ge$ 2$\rho_{a}$. The magnetic part of $\rho$ exhibits a double-peak structure with maxima at 15 K and 250 K. The magnetic entropy at $T$$\rm_{N1}$ = 3.9 K is a half of $R$ln2. These observations are ascribable to the combination of the Kondo effect with $T$$\rm_{K}$ $\sim$ 20 K and a strong crystal field effect. The analysis of $M$($B$) and paramagnetic susceptibility revealed unusually large energy splitting of 500 K and 1600 K for the two excited doublets, respectively.

\end{abstract}

\maketitle

\section{\label{sec:level1}INTRODUCTION}

\begin{figure}
\includegraphics[width=7cm]{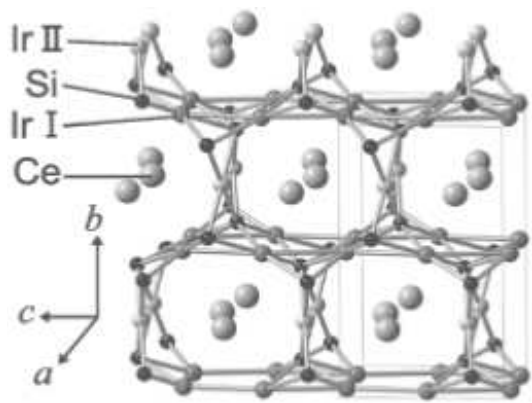}
\caption{\label{fig:Fig1.eps} Crystallographic structure of CeIr$_3$Si$_2$. Ce atoms form a zigzag chain along the $a$-axis of the orthorhombic ErRh$_3$Si$_2$-type structure. }
\end{figure}

Cerium-based intermetallic compounds crystallizing in non-cubic structures often exhibit strongly anisotropic behaviors in the magnetic and transport properties.\cite{Fujii} The anisotropy is governed by the combination of the effect of crystal electric field (CEF) on the 4$f$ electrons of Ce ions and the hybridization effect of the 4$f$ wave functions with conduction electrons.\cite{Evans} For example, strong easy-plane magnetocrystalline anisotropy appears in a hexagonal compound CeRh$_3$B$_2$ (Ref. 3) which orders ferromagnetically at an anomalously high Curie temperature of 115 K.\cite{Dhar,Malik} These properties of CeRh$_3$B$_2$ were attributed to the hybridization of the 4$f$ electron states with 5$d$ electron states of the nearest-neighbor Ce ion in a very short distance of 3.09 {\AA} along the $c$ axis of the hexagonal CeCo$_3$B$_2$-type structure.\cite{Yamaguchi} This short distance and strong hybridization may lead to a huge CEF splitting of 2000 K,\cite{Givord} which is comparable to the spin-orbit splitting between the $J$ = 5/2  ground state and the excited $J$ = 7/2 state. In fact, the contributions from the excited $J$ = 7/2 state to the 4$f$ wave functions have been confirmed by the polarized neutron scattering experiments.\cite{Givord} The CEF effect results in a strong reduction of the 4$f$ orbital magnetic moment.\cite{Schille} The nearly localized character of the 4$f$ state in CeRh$_3$B$_2$ was inferred from the fact that the 4$f$ electrons do not contribute to the Fermi surface.\cite{Okubo}  
In an isoelectronic compound CeIr$_3$B$_2$, which crystallizes in a monoclinically distorted structure of the CeCo$_3$B$_2$-type, the 4$f$ state is in the valence fluctuating regime where the localized moment is lost by the strong hybridization with conduction electrons.\cite{Yang} However, 4$f$ electrons in another related compound CeIr$_3$Si$_2$ are rather localized, as was indicated from the Curie-Weiss behavior of trivalent Ce ions in the magnetic susceptibility.\cite{Umajiri} Recent measurements of magnetic susceptibility and specific heat at low temperatures revealed magnetic orderings in CeIr$_3$Si$_2$ at $T\rm_{N1}$ = 4.1 K and $T\rm_{N2}$ = 3.3 K.\cite{Muro} Moreover, the magnetization has a spontaneous component of 0.2 $\mu_B$/Ce and exhibits three-step metamagnetic transitions.\cite{Muro} The crystal structure of CeIr$_3$Si$_2$ was found to be the orthorhombic ErRh$_3$Si$_2$-type (Space Group $Imma$, No. 74), as is illustrated in Fig.~\ref{fig:Fig1.eps}. The zigzag chain of Ce ions along the $a$-axis corresponds to the straight chain along the hexagonal $c$-axis in CeRh$_3$B$_2$. Although the Ce-Ce distance of 3.6 {\AA} in the former is longer than 3.09 {\AA} in the latter, the similar structure should lead to strong anisotropy in physical properties. With this in mind, we have grown a single crystal of CeIr$_3$Si$_2$ and herein report the magnetic, transport, and thermal properties.

\section{EXPERIMENTAL}
\subsection{Synthesis and characterization}
A single crystal of CeIr$_3$Si$_2$ was grown by the Czochralski pulling method in a tetra-arc furnace. Starting materials were Ce ingots of high purity from Ames Laboratory, shots of Ir (3N), and ingots of Si (5N). A total of about 17 g in a stoichiometric composition was melt in the furnace. A polycrystalline rod was immersed into the melt and pulled at a speed of 10 mm/h in an Ar atmosphere.\cite{Nakamoto} The grown crystal was 2-5 mm in diameter and 40 mm in length. As a reference compound with empty 4$f$ states, a polycrystalline sample of LaIr$_3$Si$_2$ was prepared by arc melting and homogenized by annealing at 900 $^{\circ}$C for 7 days. Powder X-ray diffraction was performed on both samples. The lattice parameters for CeIr$_3$Si$_2$ and LaIr$_3$Si$_2$ are $a$ = 7.178 {\AA}, $b$ = 9.726 {\AA}, $c$ = 5.597 {\AA} and $a$ = 7.189 {\AA}, $b$ = 9.763 {\AA}, $c$ = 5.615 {\AA}, respectively, whose values agree with reported ones.\cite{Muro} The single crystal was oriented by the X-ray back Laue method and cut along the principal axes for the transport and magnetic measurements.

\subsection{Measurements}

Measurements of magnetization $M$($B$) were carried out by several methods depending on the range of temperature and magnetic field. A capacitive Faraday method was employed for measurements at 0.3 K in fields up to 9 T with a field gradient of 10 T/m. To extend the field range up to 14.5 T, a sample-extraction magnetometer was used. For measurements in a wide temperature range from 1.9 K to 700 K, we used a Quantum Design MPMS. The data taken by MPMS were used to calibrate the data obtained by the Faraday force magnetometer and the sample-extraction magnetometer. The electrical resistivity $\rho$ was measured by a dc four-probe method using home-built setups from 1.3 K to 300 K. In the range up to 700 K, a closed cycle GM refrigerator was used. The specific heat was measured in magnetic fields up to 5 T and in a temperature range between 2.0 and 320 K by using a Quantum Design PPMS.

\section{RESULTS AND DISCUSSION}
\subsection{Magnetic properties}

\begin{figure}[h]
\includegraphics[width=7cm]{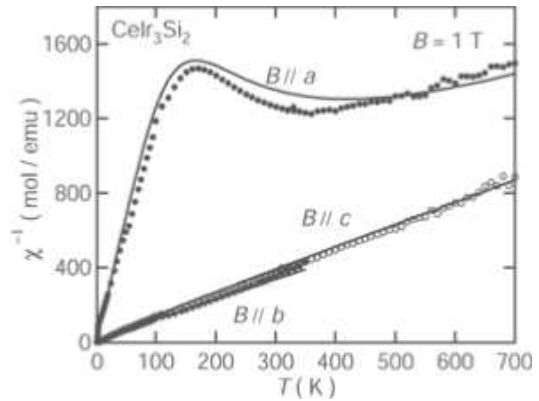}
\caption{\label{fig:Fig2.eps} Temperature dependence of the inverse magnetic susceptibility $\chi^{-1}$ $=$ $B/M$ for CeIr$_3$Si$_2$ in a field of 1 T applied along the three principal axes. The solid curves are the fits to a CEF model using Eq. (3).}
\end{figure}

Fig.~\ref{fig:Fig2.eps} shows the temperature dependence of the inverse magnetic susceptibility $B$/$M$ = $\chi$$^{-1}$ in a magnetic field $B$ = 1 T applied along the three principal axes. The data demonstrate easy-plane magnetocrystalline anisotropy with respect to the Ce chain along the $a$ axis. At 300 K, the ratio of $\chi_{B\bot a }$ / $\chi_{B//a}$ is 3, which is larger than the ratio of $\chi_{B\bot c }$ / $\chi_{B//c}$ = 2 reported for CeRh$_3$B$_2$.\cite{Kasaya2} This fact suggests the presence of strong CEF effect on the trivalent Ce ions in CeIr$_3$Si$_2$. The slopes of $\chi$$^{-1}$($T$) for $T$ $\ge$ 100 K give the effective magnetic moments $\mu_{eff}$ of 2.60 and 2.63 $\mu_B$/Ce for $B//b$ and for $B//c$, respectively. These values are close to that expected for a free trivalent Ce ion (2.54 $\mu_B$). The paramagnetic Curie temperatures are 0.8 K and -14 K for $B//b$ and $B//c$, respectively.

\begin{figure}[h]
\includegraphics[width=7cm]{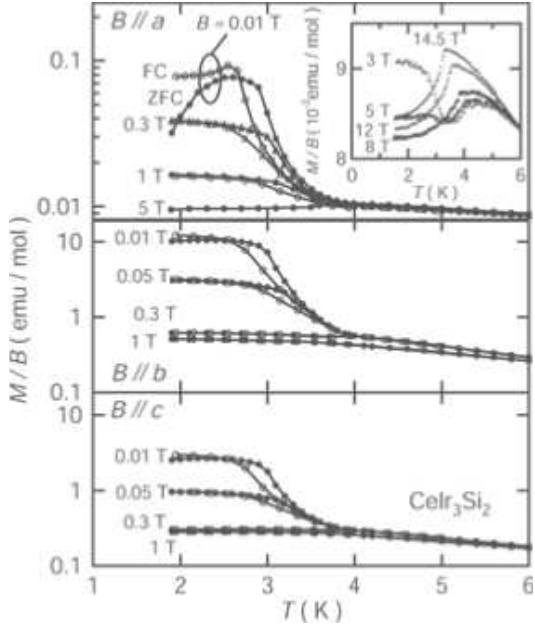}
\caption{\label{fig:Fig3.eps} Magnetization divided by external magnetic field $M/B$ for CeIr$_3$Si$_2$ at low temperatures in various magnetic fields applied along the three principal axes. Closed and open symbols represent the data in zero-field-cooling and field-cooling processes, respectively. Note the scale for $\chi_a$ is one hundredth of those for others. The inset shows $M/B$ vs $T$ in various fields $B//a$ up to 14.5 T.}
\end{figure}

At temperatures below 6 K, data of $M/B$ were recorded in both zero-field cooling (ZFC) and field cooling (FC) processes. The data are plotted in Fig.~\ref{fig:Fig3.eps}, where the vertical scale for $B//a$ is one hundredth of those for $B//b$ and $B//c$. Note that the values for $B//b$ and $B//c$ are approximately 100 times larger than for $B//a$ in a field of 0.01 T and at 2 K. The data below 4 K diverge between ZFC and FC processes. This divergence is indicative of the presence of a ferromagnetic component. The inset of Fig.~\ref{fig:Fig3.eps} is an expanded view of $M/B$ data in various fields applied along the $a$-axis. The peak at 4 K ($\cong$ $T\rm_{N1}$) for $B$ = 3 T shifts to lower temperature with increasing magnetic fields up to 14.5 T. This field dependence is a characteristic of the antiferromagnetic ordering. On the other hand, the upturn of $M/B$ below 3 K ($\cong$ $T\rm_{N2}$) for $B$ = 3 T is suppressed when the fieled is increased up to 5 T.

\begin{figure}[h]
\includegraphics[width=7cm]{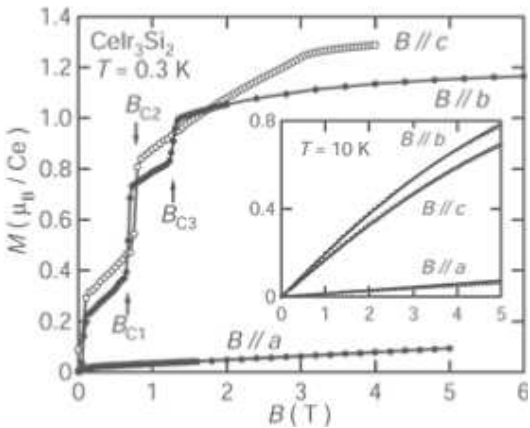}
\caption{\label{fig:Fig4.eps} Magnetization processes $M$($B$) of CeIr$_3$Si$_2$ in magnetic fields along the three principal axes. The data for $B//b$ and $B//c$ are taken at 0.3 K, while data for $B//a$ are at 1.9 K. The inset shows $M$($B$) at 10 K in the paramagnetic state. The solid curves are the fits to the data with a CEF model (see text).}
\end{figure}

The easy-plane magneto-crystalline anisotropy in CeIr$_3$Si$_2$ is highlighted in the isothermal magnetization curves $M$($B$) at 0.3 K as is shown in Fig.~\ref{fig:Fig4.eps}. The relation $M_b$ $\cong$ $M_c$ $\cong$ 11 $M_a$ at $B$ = 5 T indicates that the $a$-axis is the hard magnetization direction. It should be noted that this anisotropy in CeIr$_3$Si$_2$ is stronger than that in CeRh$_3$B$_2$ where $M_{B\bot c }$ is four times of $M_{B//c}$ at 7 T.\cite{Galatanu} In Fig.~\ref{fig:Fig4.eps}, there are two metamagnetic transitions at $B\rm_{C1}$ = 0.68 T and $B\rm_{C3}$ = 1.3 T for $B//b$, while for $B//c$ the former shifts to $B\rm_{C2}$ = 0.75 T and a kink appears at 3 T. No hysteresis was observed between increasing and decreasing runs. The values for $B\rm_{C1}$, $B\rm_{C2}$, and $B\rm_{C3}$ agree with those reported for a polycrystalline sample.\cite{Muro} In higher fields up to 14.5 T, $M$($B$) was measured by the extraction method. It was observed that $M$($B$) at 1.5 K linearly increases to values of 0.20, 1.23, and 1.38$\mu_B$/Ce at 14.5 T for $B//a$, $B//b$, and $B//c$, respectively. The inset of Fig.~\ref{fig:Fig4.eps} shows the magnetization curves in the paramagnetic state at 10 K.

We analyzed the two sets of data $\chi$($T$) and $M$($B$) using a CEF model, considering orthorhombic site symmetry of Ce ions in CeIr$_3$Si$_2$. According to Hutchings' notation,\cite{Hutchings} the CEF Hamiltonian for $J$ = 5/2 with the orthorhombic point symmetry is given by 
\begin{eqnarray}%
\mathcal{H}_{CEF} = B_2^0O_2^0+B_2^2O_2^2+B_4^0O_4^0+B_4^2O_4^2+B_4^4O_4^4%
\end{eqnarray}%

where $B_n^m$ and $O_n^m$ represent the CEF parameters and the Steven's equivalent operators, respectively. In the mean-filed approximation, $\chi_i$ ($T$) ($i$ $=$ $a, b, c$) is expressed as

\begin{eqnarray}%
\chi_i (T) = \chi_{iCEF} (T)/ (1-\lambda\chi_{iCEF} (T))  
\end{eqnarray}%

where the CEF susceptibility $\chi_{iCEF}$ is given by

\begin{align}
 \chi_{i \mathrm{CEF}} &= \frac{(g_J \mu_\mathrm{B})^2}{\sum\limits_m
e^{-E_m /k_\mathrm{B} T}}
 \Bigg( \frac{\sum\limits_m |\langle m | J_i | m \rangle |^2 e^{-E_m
/k_\mathrm{B} T}}{k_\mathrm{B} T} \nonumber\\
 &+ \sum_m \sum_{n(\neq m)} | \langle n | J_z | m \rangle
|^2\frac{e^{-E_m /k_\mathrm{B} T} - e^{-E_n /k_\mathrm{B} T}}{E_n - E_m}
\Bigg) 
\end{align}

The CEF magnetization $M_{iCEF}$ is given by

\begin{align}
 M_{iCEF}  &= g_J \mu_{\mathrm{B}} \sum_m |\langle m | J_i | m \rangle |
\frac{e^{-E_m / k_{\mathrm{B}} T}}{\sum\limits_m e^{-E_m /
k_{\mathrm{B}} T}}
\end{align}                                                             
where $g_J$ is the Land$\acute{\rm{e}}$ $g$-facter, $J_i$ is a component of the angular momentum, $m$ and $n$ are eigenstates of 4$f$ wave functions, and $E_{(m,n)}$ is the eigenvalue of the derived CEF energy level.
The CEF parameters were determined by a least-square fitting to the data of $\chi$$^{-1}$($T$) and $M$($B$) in the paramagnetic region. Thereby, the quantization axis was chosen as the $b$ axis, because the paramagnetic susceptibility is largest for $B//b$. The best fit was obtained with the parameters listed in Tables~\ref{tab:table1}. The calculated curves are drawn by solid lines in Fig.~\ref{fig:Fig2.eps} and the inset of Fig.~\ref{fig:Fig4.eps}, respectively. The fairly good fitting for both sets of the data, especially the pronounced maximum in $\chi$$^{-1}$($T$) for $B//a$, suggests that the strong anisotropy in CeIr$_3$Si$_2$ can be understood by the CEF effects. It is noteworthy that largely negative values for the higher order terms $B^2_4$ and $B^4_4$ are necessary to reproduce the maximum in $\chi^{-1}$ for $B//a$. The CEF parameters lead to a crystal field level scheme of three doublets at 0 K, 500 K ($\Delta_1$), 1620 K ($\Delta_2$), respectively. This huge CEF splitting is comparable to that of CeRh$_3$B$_2$ (2000 K).\cite{Givord} The wave function of the ground state consists mainly of $\ket{\pm5/2}$, whose function elongates along the $a$ axis. In this configuration, the magnetic moment should orient perpendicular to the $a$ axis. This idea is supported by the experimental fact that the hard magnetic direction is along the $a$ axis.

\begin{table*}
\caption{\label{tab:table1}Crystal field parameters, molecular field coefficient, energy levels, and wave functions for CeIr$_3$Si$_2$ obtained from the fitting of magnetic susceptibility and magnetization in the paramagnetic state.}
\begin{ruledtabular}
\begin{tabular}{ccccccc}
 \multicolumn{7}{c}{\rm{CEF parameters}}\\ \
  &$B_2^0$ (K)&$B_2^2$&$B_4^0$&$B_4^2$&$B_4^4$&$\lambda$ $(emu/mol)^{-1}$\\ 
   &-24.5&-39.7&1.03&-18.24&-14.5&5.92\\ \hline
 Energy levels& & &Wave functions& & &\\ \hline
 $E$ (K)&$\ket{+5/2}$&$\ket{+3/2}$&$\ket{+1/2}$&$\ket{-1/2}$&$\ket{-3/2}$&$\ket{-5/2}$\\
1620&0.564&0&-0.742&0&-0.363&0\\
1620&0&0.363&0&0.742&0&-0.564\\
500&0&0.847&0&-0.515&0&-0.133\\
500&0.133&0&0.515&0&-0.847&0\\
0&0&0.389&0&0.429&0&0.815\\
0&-0.815&0&-0.429&0&-0.389&0\\
 
\end{tabular}
\end{ruledtabular}
\end{table*}

\subsection{Transport properties}

  \begin{figure}[h]
\includegraphics[width=7cm]{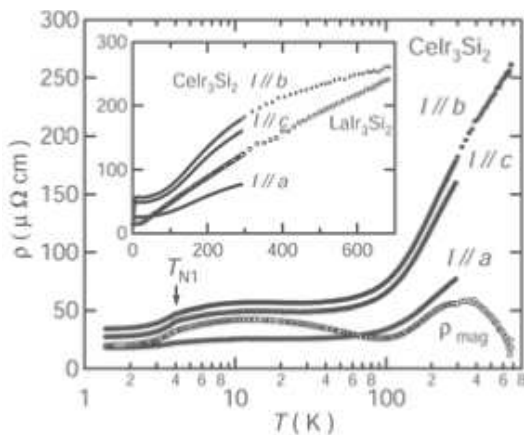}
\caption{\label{fig:Fig5.eps} Electrical resistivity of CeIr$_3$Si$_2$ as a function of temperature for the current directions $I//a$, $I//b$ and $I//c$. The magnetic contribution $\rho\rm_{mag}$ ($I//b$) was derived by the subtraction of the resistivity of a polycrystalline LaIr$_3$Si$_2$ shown in the inset. }
\end{figure}

The electrical resistivity $\rho$ along the three principal axes is plotted versus log$T$ and $T$ in Fig.~\ref{fig:Fig5.eps} and the inset, respectively. The relation $\rho$$_c$ $\cong$ $\rho$$_b$ $\ge$ 2$\rho$$_a$ holds in the whole temperature range. The low resistivity along the $a$ axis may reflect the electronic conduction through the honeycomb-like network of Ir and Si atoms parallel to the Ce-Ce zigzag chain (see Fig. 1). The lower resistivity parallel to the chain is common to the series of compounds RRh$_3$B$_2$ (R $=$ La, Ce, Pr and Gd).\cite{Yamada} A sudden drop in $\rho$($T$) at $T$$\rm_{N1}$ = 3.9 K is ascribable to the reduction of the scattering of conduction electrons
 due to the partial alignment of Ce magnetic moments. At $T$$\rm_{N2}$ = 3.1 K, where the specific heat exhibits a peak as will be shown later, the differential resistivity d$\rho$/d$T$ shows a maximum. The magnetic contribution $\rho$$\rm_{mag}$ along the $b$ axis was estimated by subtracting the resistivity of LaIr$_3$Si$_2$ without 4$f$ electrons; $\rho_b$(CeIr$_3$Si$_2$)|$\rho$(LaIr$_3$Si$_2$). Thereby, no anisotropy in the La compound was assumed. The plot of $\rho_m$ vs ln$T$ in Fig.~\ref{fig:Fig5.eps} shows the $-$ln$T$ behavior in two temperature ranges from 30 K to 70 K and from 400 K to 700 K, respectively. If we take the temperature where $\rho$$\rm_{mag}$($T$) swerves from the $-$ln$T$ dependence as the Kondo temperature, then $T$$\rm_K$ is approximately 20 K. Another measure of the $T$$\rm_K$ on the whole multiplet including the excited CEF levels is given by $T$$\rm_K^h$ = ($T$$\rm_K$$\Delta_1$$\Delta_2$)$^{1/3}$.\cite{Hanzawa} Using $T$$\rm_K$ = 20 K, $\Delta_1$ = 500 K, and $\Delta_2$ = 1620 K, $T$$\rm_K^h$ is determined as 250 K, which is close to the temperature where the upper maximum appears in $\rho$$\rm_{mag}$($T$).

\begin{figure}[h]
\includegraphics[width=7cm]{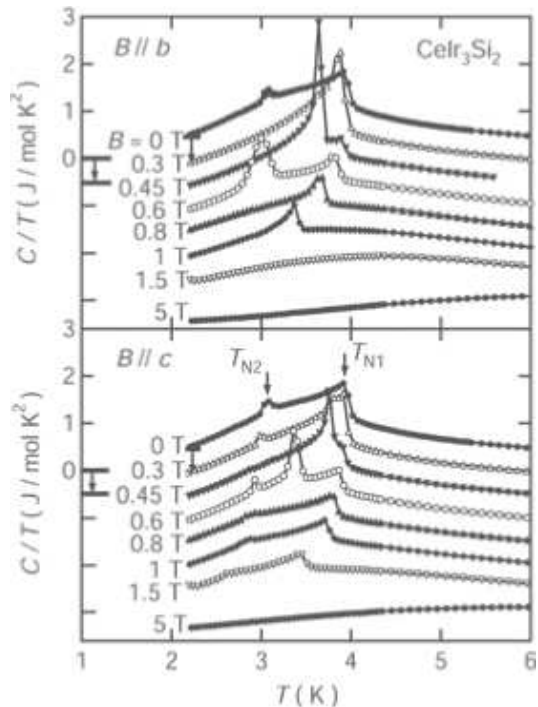}
\caption{\label{fig:Fig6.eps}  Specific heat divided by temperature $C/T$ vs $T$ for CeIr$_3$Si$_2$ in various magnetic fields applied parallel to the $b$-and $c$-axes. The origin of the ordinate is shifted downward for each set of data.}
\end{figure}

 To study the nature of the double magnetic transitions of CeIr$_3$Si$_2$, the specific heat was measured in various magnetic fields up to 5 T. The $C$/$T$ vs $T$ plots for $B//b$ and $B//c$ are shown in Fig.~\ref{fig:Fig6.eps}. In zero field, two distinct peaks exist at $T$$\rm_{N1}$ = 3.9 K and $T$$\rm_{N2}$ = 3.1 K, respectively. When the magnetic field is applied, the two anomalies in $C/T$ shift to lower temperature. Above 0.3 T, the peak at $T$$\rm_{N1}$ splits into two anomalies at $T$$\rm_{N1}$ and $T$$\rm_{N1}^*$. The weak peak at $T$$\rm_{N2}$, on the other hand, moves to below 2 K already at 0.3 T for $B//b$, while it stays above 2 K even at 1 T for $B//c$.

\begin{figure}
\includegraphics[width=7cm]{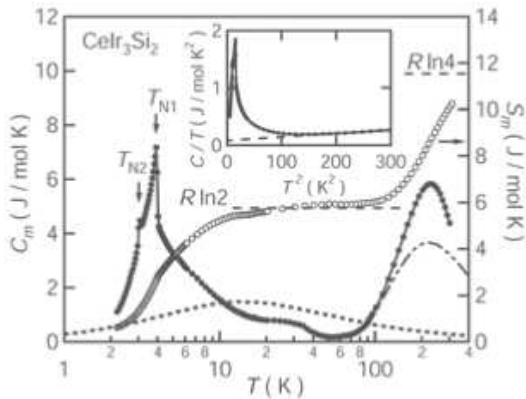}
\caption{\label{fig:Fig7.eps} Temperature dependence of the magnetic part of specific heat $C\rm_m$ and the magnetic entropy $S\rm_m$ for CeIr$_3$Si$_2$. The dotted line represents the theoretical curve of $C\rm_m$ for $S$ = 1/2 Kondo impurities with $T\rm_K$ = 20 K, and the dashed-dotted line is the crystal-field contributions to $C\rm_m$ calculated for $\Delta$ = 530 K. The inset is the plot of $C/T$ vs $T^2$.}
\end{figure}

   The data of $C/T$ are plotted vs $T^2$ in the inset of Fig.~\ref{fig:Fig7.eps}. A linear extrapolation of the data above 13 K to $T$ = 0 yields the electronic specific heat coefficient $\gamma$ = 94 mJ/molK$^2$. This $\gamma$ value is translated to $T\rm_K$ =20 K by using the relation $\gamma$ (mJ/molK$^2$) = 5627/3$T\rm_K$(K), which is relevant for a doublet ground state in the Coqblin-Schrieffer model.\cite{Rajan} This value of $T\rm_K$ agrees with that estimated from the -ln$T$ dependence of $\rho\rm_{mag}$. The magnetic contribution to the specific heat, $C\rm_m$, is derived by subtracting the value of $C$ for LaIr$_3$Si$_2$ from that of CeIr$_3$Si$_2$. The temperature variation of $C\rm_m$ shown in Fig.~\ref{fig:Fig7.eps} has a significant tail in a wide range from $T\rm_{N1}$ = 3.9 K to 10 K. The tail can not be reproduced by the Kondo contribution with $T\rm_{K}$ = 20 K, which is drawn by a dotted curve. The tail is therefore ascribed to the development of antiferromagnetic short-range correlations along the $a$-axial before the long-range order sets in. The maximum at $T_{C\rm max}$ = 220 K is a Schottky anomaly due to CEF excitation from the ground state to the first excited doublet. An energy splitting of $\Delta_1$ = 530 K is estimated as 2.4 times of $T_{C\rm max}$, whose value is in good agreement with the result obtained by CEF analysis. The magnetic entropy $S\rm_{m}$ at $T\rm_{N1}$ = 3.9 K is only 0.5$R$ln2, therefore greatly reduced by the entropy consumption due to the Kondo effect and short-range correlations. With increasing temperature, $S\rm_{m}$ recovers the full value of $R$ln2 at 30 K.

\begin{figure}
\includegraphics[width=7cm]{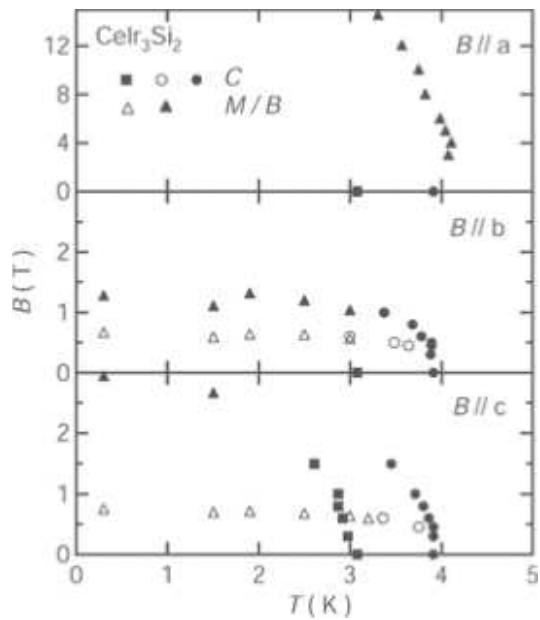}
\caption{\label{fig:Fig8.eps} A proposed $B$-$T$ phase diagram of CeIr$_3$Si$_2$ for $B//a$, $B//b$ and $B//c$. The circles and squares denote the data taken from the specific heat anomaly and the triangles are from the metamagnetic transitions.}
\end{figure}

In an attempt to construct the magnetic phase diagram of CeIr$_3$Si$_2$, we plot in Fig.~\ref{fig:Fig8.eps} the values of the field and temperature of the specific-heat anomalies together with the fields of metamagnetic transitions. For the hard direction, $B//a$, the phase boundary between the paramagnetic state and the antiferromagnetic state gradually decreases with increasing field. For $B//b$ and $B//c$, the antiferromagnetic phase is quenched at a rather low field of a few T. The phase boundary lying at 0.5 T may be the spin-flop transition. 

\section{Summary}

 Magnetic and transport measurements on a single crystal of CeIr$_3$Si$_2$, which crystallizes in the orthorhombic ErRh$_3$Si$_2$-type structure, revealed highly anisotropic properties. The high-temperature properties can be understood in the framework of perturbation theory including the CEF effects on trivalent Ce ions. The overall CEF splitting of 1600 K is anomalously large among Ce compounds.\cite{Moze} It is comparable with that in the well-studied ferromagnet CeRh$_3$B$_2$ with a closely related crystal structure. In both compounds, chains of Ce ions run in a honeycomb-like network. Photoemission study is necessary to investigate whether the giant CEF splitting in CeIr$_3$Si$_2$ originates from the hybridization of 4$f$ electrons with conduction bands. It is also interesting to examine whether the higher $J$=7/2 multiplet plays a role in the anisotropic properties as was found in CeRh$_3$B$_2$. The first CEF excited level at 500 K in CeIr$_3$Si$_2$ manifests itself as the Schottky anomaly in the specific heat and the -log$T$ behavior in the magnetic part of the electrical resistivity. At low temperatures, the RKKY-type interaction among the magnetic moments of the CEF ground state overwhelms the Kondo effect, leading to antiferromagnetic order at 3.9 K. Below the second transition at 3.1 K, a ferromagnetic component appears along the $b$ and $c$ axes. When magnetic field along these directions is increased, metamagnetic transitions occur at rather low fields. At $B$ = 5 T, both $M$($B//b$) and $M$($B//c$) are one order of magnitude larger than $M$($B//a$). In order to understand the origin of this strong anisotropy and complex magnetic phase diagrams, neutron scattering experiments are in progress.

\begin{acknowledgments}
 We thank Y. Ando for his help in crystal field analysis of magnetic susceptibility, R. Morinaga and T. J. Sato for the generous use of the high-temperature SQUID magnetometer, T. Sakakibara for the valuable advice for the design of capacitive Faraday magnetometer. We are thankful to F. Iga, M.A. Avila and T. Saso for fruitful discussions. This work was financially supported by the Grant-in-Aids for Scientific Research (A), No. 18204032 and the priority area gSkutteruditeh No. 15072205 from MEXT, Japan. 
\end{acknowledgments}

\end{document}